\documentclass[12pt]{article}
\usepackage{graphics,psfrag}
\usepackage{amsmath,amsthm,amssymb,epsfig,euscript,array,cite,cancel,color}
\usepackage{cases,empheq}
\usepackage[colorlinks=true]{hyperref}
\usepackage{verbatim}
\usepackage{url}
\usepackage{bbm}
\usepackage{mciteplus}
   \usepackage[vcentermath]{youngtab}  
   
   \usepackage[usenames,dvipsnames]{xcolor}

\newcounter{multieqs}

\newcommand{\be}{\begin{equation}}
\newcommand{\ee}{\end{equation}}
\newcommand{\eq}[1]{(\ref{#1})}
\newcommand{\bit}{\begin{itemize}}  \newcommand{\eit}{\end{itemize}}

\newcommand{\bm}[1]{\mbox{\boldmath $#1$}}
\newcommand{\rf}[1]{(\ref{#1})}

\def\bd{\begin{document}}
\def\ed{\end{document}}
\def\nn{\nonumber}
\def\bea{\begin{eqnarray}}
\def\eea{\end{eqnarray}}
\let\bm=\bibitem

\def\la{\langle}
\def\ra{\rangle}

\def\N{{\cal N}}
\def\sst{\scriptscriptstyle}
\def\thetabar{\bar\theta}
\def\Tr{{\rm Tr}}
\def\one{\mbox{1 \kern-.59em {\rm l}}}

\def\a{\alpha}      \def\da{{\dot\alpha}}  \def\dA{{\dot A}}
\def\b{\beta}       \def\db{{\dot\beta}}  
\def\g{\gamma}  \def\G{\Gamma}  \def\dc{{\dot\gamma}}  
\def\d{\delta}  \def\D{\Delta}  \def\ddt{\dot\delta}  
\def\e{\epsilon}        \def\ve{\varepsilon}  
\def\f{\phi}    \def\F{\Phi}    \def\vvf{\f}  
\def\h{\eta}  
\def\k{\kappa}  
\def\l{{\lambda}} \def\L{\Lambda}  
\def\m{\mu} \def\n{\nu}  
\def\o{\omega}  
\def\p{\pi} \def\P{\Pi}  
\def\r{\rho}  
\def\s{\sigma}  \def\S{\Sigma}  
\def\t{\tau}  
\def\th{\theta} \def\Th{\Theta} \def\vth{\vartheta}  
\def\X{\Xeta}  
\def\z{\zeta}  

\def\na{\nabla}  

\def\cA{{\cal A}} \def\cB{{\cal B}} \def\cC{{\cal C}}  
\def\cD{{\cal D}} \def\cE{{\cal E}} \def\cF{{\cal F}}  
\def\cG{{\cal G}} \def\cH{{\cal H}} \def\cI{{\cal I}}  
\def\cJ{{\cal J}} \def\cK{{\cal K}} \def\cL{{\cal L}}  
\def\cM{{\cal M}} \def\cN{{\cal N}} \def\cO{{\cal O}}  
\def\cP{{\cal P}} \def\cQ{{\cal Q}} \def\cR{{\cal R}}  
\def\cS{{\cal S}} \def\cT{{\cal T}} \def\cU{{\cal U}}  
\def\cV{{\cal V}} \def\cW{{\cal W}} \def\cX{{\cal X}}  
\def\cY{{\cal Y}} \def\cZ{{\cal Z}}

\def\ua{{\underline{\alpha}}} 
 \def\ub{\underline{\phantom{\alpha}}\!\!\!\beta}  
\def\uc{\underline{\phantom{\alpha}}\!\!\!\gamma}  
\def\um {{\underline{\mu}}}  
\def\ud{{\underline{\delta}}} 
\def\ue{\underline\epsilon}  
\def\una{{\underline a}}\def\uA{{\underline A}}  
\def\unb{{\underline b}}\def\uB{{\underline B}} 
\def\unc{{\underline c}}\def\uC{{\underline C}}  
\def\und{{\underline d}}\def\uD{{\underline D}}  
\def\une{{\underline e}}\def\uE{{\underline E}}  
\def\unf{{\underline{\phantom{e}}\!\!\!\! f}}\def\uF{{\underline F}}  
\def\unm{{\underline m}\def\uM{\underline M}} 
\def\unn{{\underline n}\def\uN{\underline N}} 
\def\unp{{\underline{\phantom{a}}\!\!\! p}}\def\uP{{\underline P}}  
\def\unq{{\underline{\phantom{a}}\!\!\! q}}  
\def\uQ{{\underline{\phantom{A}}\!\!\!\! Q}}  
\def\uH{{\underline{H}}}  
\def\uM{{\underline{M}}}
\def\uN{{\underline{N}}}
\def\unl{{\underline{l}}}
  
\def\As {{A \hspace{-6.4pt} \slash}\;}  
\def\bs {{b \hspace{-6.4pt} \slash}\;}  
\def\Ds {{D \hspace{-6.4pt} \slash}\;}
\def\Gts {{\Gt \hspace{-6.4pt} \slash}\;}
\def\ds {{\del \hspace{-6.4pt} \slash}\;}  
\def\ss {{\s \hspace{-6.4pt} \slash}\;}  
\def\ks {{ k \hspace{-6.4pt} \slash}\;}  
\def\ps {{p \hspace{-6.4pt} \slash}\;}   
\def\xs {{x \hspace{-6.4pt} \slash}\;}  
\def\pas {{{p_1} \hspace{-6.4pt} \slash}\;}  
\def\pbs {{{p_2} \hspace{-6.4pt} \slash}\;}   
\def\cFs {{{\cal F} \hspace{-6.4pt} \slash}\;}

\def\Ah{{\hat{A}}}  
\def\Dh{{\hat{D}}}
\def\Gh{{\hat{G}}}
\def\Fh{{\hat{F}}}
\def\Ih{{\hat{I}}} 
\def\Jh{{\hat{J}}} 
\def\Kh{{\hat{K}}}
\def\Lh{{\hat{L}}} 
\def\Ph{{\hat{P}}}
\def\Rh{{\hat{R}}}
\def\Vh{{\hat{V}}} 
\def\Xh{{\hat{X}}}
 
\def\ah{{\hat{a}}}
\def\bh{{\hat{b}}}
\def\ch{{\hat{c}}}
\def\gh{{\hat{g}}}
\def\dh{{\hat{d}}}
\def\hh{{\hat{h}}}
\def\uh{{\hat{u}}}  
\def\vh{{\hat{v}}}
\def\xh{{\hat{x}}} 
\def\yh{{\hat{y}}}
\def\zh{{\hat{z}}}
\def\ph{{\hat{p}}}
\def\qh{{\hat{q}}}
\def\thh{{\hat{t}}}  
\def\xih{\hat{\xi}}  
\def\Psih{\hat{\Psi}}    
\def\mh{{\hat{m}}}
\def\nh{{\hat{n}}}
\def\ih{{\hat{i}}}
\def\jh{{\hat{j}}}
\def\kh{{\hat{k}}}
\def\aah{{\hat{\alpha}}}
\def\bbh{{\hat{\beta}}}
\def\ggh{{\hat{\gamma}}}
\def\llh{{\hat{\ell}}} 
\def\ph{{\hat{p}}}

\def\psit{\tilde{\psi}}  
\def\Psit{\tilde{\Psi}}   
\def\Psibt{\tilde{\bar{Psi}}}  

\def\st{\tilde{\sigma}}  

\def\delt{\tilde{\delta}}
\def\Phit{\tilde{\Phi}}   
\def\Phitb{\overline{\tilde{Phi}}}  
\def\tht{\tilde{\th}}  
\def\lt{\tilde{\l}}
\def\chit{\tilde{\chi}}   
\def\phit{\tilde{\phi}} 

\def\At{\tilde{A}}
\def\Bt{\tilde{B}}
\def\Ct{\tilde{C}}
\def\Dt{\tilde{D}}
\def\Et{\tilde{E}}
\def\Ft{\tilde{F}}
\def\Gt{\tilde{G}}
\def\Ht{\tilde{H}}
\def\It{\tilde{I}}
\def\Jt{\tilde{J}}
\def\Qt{\tilde{Q}}  
\def\Rt{\tilde{R}}  
\def\Mt{\tilde{M }}  
\def\Nt{\tilde{N}}   
\def\St{\tilde{S}}
\def\Vt{\tilde{V}}
\def\Xt{\tilde{X}} 
\def\at{\tilde{a}}
\def\ct{\tilde{c}}
\def\dt{\tilde{d}}
\def\htt{\tilde{h}} 
\def\ft{\tilde{f}}
\def\gt{\tilde{g}}
\def\pt{\tilde{p}}  
\def\qt{\tilde{q}}  
\def\vt{\tilde{v}}  
\def\nt{\tilde{n}}  
\def\ut{\tilde{u}}  
\def\wt{\tilde{w}}  
\def\zt{\tilde{z}} 
\def\xt{\tilde{x}} 
\def\yt{\tilde{y}} 
\def\Psit{\tilde{\Psi}}
\def\vphit{\tilde{\varphi}}

\def\gamt{\tilde{\gamma}}

\def\Tt{\tilde{T}}

\def\eb{\bar{\epsilon}} 
\def\delb{\bar{\partial}}  
\def\thb{\bar{\theta}}
\def\Thb{{\bar{\Theta}}}
\def\mub{\bar{\mu}}
\def\lamb{\bar{\l}}
\def\psib{\bar{\psi}}
\def\sb{\bar{\sigma}}
\def\xib{\bar{\xi}}
\def\chib{\bar{\chi}}

\def\Psib{\bar{\Psi}}
\def\Phib{\bar{\Phi}}
\def\Lamb{\bar{\Lambda}}
\def\Sb{{\overline \Sigma}}
\def\cb{\bar{c}}
\def\hb{\bar{h}}
\def\qb{\bar{q}}
\def\wb{\bar{w}}
\def\zb{{\bar{z}}}
\def\Hb{\bar{H}}
\def\Qb{{\bar Q}}
\def\Omegab{\overline{\Omega}}
\def\ob{\overline{\omega}}
\def\Gab{{\bar{\Gamma}}}

\def\Ab{{\overline A}} \def\Bb{{\overline B}} \def\Cb{{\overline C}}  
\def\Db{{\overline D}} \def\Eb{{\overline E}} \def\Fb{{\overline F}}  
\def\Gb{{\overline G}} 
\def\Ib{{\overline I}}  
\def\Jb{{\overline J}} \def\Kb{{\overline K}} \def\Lb{{\overline L}}  
\def\Mb{{\overline M}} \def\Nb{{\overline N}} \def\Ob{{\overline O}}  
\def\Pb{{\overline P}}  \def\Rb{{\overline R}}  
 \def\Tb{{\overline T}} \def\Ub{{\overline U}}  
\def\Vb{{\overline V}} \def\Wb{{\overline W}} \def\Xb{{\overline X}}  
\def\Yb{{\overline Y}} \def\Zb{{\overline Z}}  

\def\fb{{\overline f}}
\def\gb{{\overline g}}
\def\mb{{\overline m}}
\def\lb{{\overline l}}
\def\yb{{\overline y}}
  
\def\ldel{{\overleftarrow{\del}}}
\def\rdel{{\overrightarrow{\del}}}
\def\ldeldel{{\overleftarrow{\del^2}}}
\def\rdeldel{{\overrightarrow{\del^2}}}
\def\ldelb{{\overleftarrow{\bar{\del}}}}
\def\rdelb{{\overrightarrow{\bar{\del}}}}

\def\ba{{\bf a}} 
\def\bk{{\bf k}}  
\def\bl{{\bf l}}  
\def\bp{{\bf p}}  
\def\bq{{\bf q}}  
\def\br{{\bf r}}
\def\bt{{\bf t}}
\def\bu{{\bf u}}
\def\bv{{\bf v}}
\def\bx{{\bf x}}  
\def\by{{\bf y}}  
\def\bR{{\bf R}}  
\def\bV{{\bf V}}

\def\bone{{\bf 1}}  

\def\va{{\vec a}}
\def\vk{{\vec k}}
\def\vp{{\vec p}}
\def\vq{{\vec q}}
\def\vx{{\vec x}}
\def\vy{{\vec y}}
\def\vu{{\vec u}}
\def\vv{{\vec v}}

\def\vs{{\vec \sigma}}
\def\vtau{{\vec \tau}}

\newcommand{\ov}[1]{\overrightarrow{#1}}

\def\d{\delta}\def\D{\Delta}\def\ddt{\dot\delta}  
  
\def\pa{\partial} \def\del{\partial}  
\def\xx{\times}  
\def\uno{\mbox{1 \kern-.59em {\rm l}}}    
  
\def\trp{^{\top}}  
\def\inv{^{-1}}  
\def\dag{{^{\dagger}}}  
\def\pr{^{\prime}}  
  
\def\rar{\rightarrow}  
\def\lar{\leftarrow}  
\def\lrar{\leftrightarrow}  
  
\newcommand{\0}{\,\!}      
\def\one{1\!\!1\,\,}  
\def\im{\imath}  
\def\jm{\jmath}  
  
\newcommand{\tr}{\mbox{tr}}  
\newcommand{\slsh}[1]{/ \!\!\!\! #1}  
  
\def\vac{|0\rangle}  
\def\lvac{\langle 0|}  
  
\def\hlf{\frac{1}{2}}  
\def\ove#1{\frac{1}{#1}}  

\def\Box{\square}  
\def\CC {\mathbb{C}}
\def\FF {\mathbb{F}}
\def\RR{\mathbb{R}}
\def\NN{\mathbb{N}}  
\def\ZZ{\mathbb{Z}}  
\def\bb#1{{\bf #1}}  
\def\bcomment#1{}  
\def\bfhat#1{{\bf \hat{#1}}}  
\def\VEV#1{\left\langle #1\right\rangle}  

\newcommand{\ex}[1]{{\rm e}^{#1}} \def\ii{{\rm i}}  

\newcommand{\lrbrk}[1]{\left(#1\right)}
\newcommand{\lrsbrk}[1]{\left[#1\right]}
\newcommand{\lrcbrk}[1]{\left\{#1\right\}}
\newcommand{\sfrac}[2]{{\textstyle\frac{#1}{#2}}}
 
\def\stw{{\sqrt{2}}}

\def\rf {{\rm f}}
\def\ri {{\rm i}}
\def\rj {{\rm j}}
\def\rk {{\rm k}}
\def\rl {{\rm l}}
\def\rs {{\scriptscriptstyle \rm S}}
\def\rt {{\scriptscriptstyle \rm T}}

\def\rQ {{\scriptscriptstyle \rm \cQ}}
\def\rR {{\scriptscriptstyle \rm \cR}}

\def\cQb{{\cal \Qb}}
\def\cRb{{\cal \Rb}}
\def\cWb{{\cal \Wb}}

\def\fd {{\rm N}}
\def\afd {{\overline{\rm N}}}

\def \II {I\hspace{-.1em}I\hspace{.1em}}
\def \IIA {\mbox{\II A\hspace{.2em}}}
\def \IIB {\mbox{\II B\hspace{.2em}}}
\def \gs {g^s}
\def \ls {\lambda^s}

\def \I {{\cal I}}
\def \qs {q\hspace{-.53em}/\hspace{.15em}}
\def \ks {k\hspace{-.53em}/\hspace{.15em}}
\def \YM {{\mbox{\tiny YM}}}
\def \gym {g_{\YM}}

\def \Lc {\L_c}
\def\IR{\relax{\rm I\kern-.18em R}}
\def \id {{\bf 1}}

\def\cci{\ell}
\def\ccj{\ell'}

\def \thbb{\overline{\th\th}}
\newcommand \ol{\overline}
\def \lamb{\bar{\lambda}}
\def \vphi{\varphi}
\def \lambh{\hat{\bar{\lambda}}}
\def \lh{\hat{\lambda}}
\def \dd{\ddagger}

\def \ad {\dot{a}}
\def \bd {\dot{b}}
\def \cd {\dot{c}}
\def  \ddd {\dot{d}}
\def \ed {\dot{e}}
\def \fd {\dot{f}}
\def \Bh {\hat{B}}
\def \zm {{(0)}}
\def \nz {{(\text{KK})}}
\def \3{{(3)}}
\def \diag {\text{diag}}
\def \inm {{(m^{-1})}}
\def \3{{(3)}} 
\def \6{{(6)}}
\def \2{{(2)}}
\def \7{{(7)}} 
\def \4{{(4)}}
\def\1{{(1)}}
\def\5{{(5)}}
\def\0{{(0)}}

\def\eh{{\hat{e}}}
\def\fh{{\hat{f}}}
\def\lh{{\hat{l}}}
\def\rh{{\hat{r}}}
\def\wh{{\hat{w}}}
\renewcommand{\mh}{{\hat{m}}}

\def \DBI{{\text{DBI}}}
\def\et{{\tilde{\e}}}
\def\w{{\wedge}}
\def\bbV{{\mathbb{V}}}
\def\M{{(\text{M})}}
\def\T{{(\text{T})}}

\def\Hbt{{\tilde{\bar{H}}}}
\def\Fbt{{\tilde{\bar{F}}}}

\colorlet{1}{red}
\colorlet{2}{green}
\colorlet{3}{blue}
\colorlet{4}{cyan}

\def\PI{{P^\perp}}
\def\Pt{{\tilde{P}}}
\def\PIt{{\tilde{P}^\perp}}
\def\cPI{{\cP^\perp}}
\def\cPt{{\tilde{\cP}}}
\def\cPIt{{\tilde{\cP}^\perp}}
\def\cun{\qquad {\color{blue}\textrm{(cun)}}}
\def\cmod{\qquad {\color{blue}\textrm{(cmod)}}}
\def\cna{\qquad {\color{blue}\textrm{(cna)}}}
\def\culater{\qquad {\color{blue}\textrm{(culater)}}}
\def\bigw{{\bigwedge}}

\def\rb{{\bar{r}}}
\def\SS{{\hlf\dot{S}}}

\author{Pichet Vanichchapongjaroen\footnote{pichetv@nu.ac.th}$~$
\\
\\
{\small  
	\it The Institute for Fundamental Study ``The Tah Poe Academia Institute",}
\\
{\small\it Naresuan University, Phitsanulok 65000, Thailand}}

\title{\bf Dual formulation of covariant nonlinear duality-symmetric action of kappa-symmetric D3-brane}

\begin{document}
\maketitle

\abstract{
We study the construction of covariant nonlinear duality-symmetric actions
in dual formulation. Essentially, the construction is the PST-covariantisation
and nonlinearisation of Zwanziger action.
The covariantisation made use of three auxiliary scalar fields.
Apart from these, the construction proceed in a similar way to that of the standard formulation.
For example, the theories can be extended to include interactions with
external fields, and that the theories possess two local PST symmetries.
We then explicitly demonstrate the construction of
covariant nonlinear duality-symmetric actions
in dual formulation of DBI theory, and D3-brane.
For each of these theories, the twisted self-duality condition
obtained from duality-symmetric actions
are explicitly shown to match with the duality relation
between field strength and its dual from the one-potential actions.
Their on-shell actions between the duality-symmetric and the one-potential versions
are also shown to match.
We also explicitly prove kappa-symmetry of the 
covariant nonlinear duality-symmetric  D3-brane action
in dual formulation.
}

\thispagestyle{empty}
\newpage
\tableofcontents

\section{Introduction}
    Duality symmetry has been an important symmetry occurring in supergravity and string theory.
  A prototypical model containing this symmetry is classical electrodynamics.
  Maxwell's equations in free space is invariant under the rotation of
  the doublet $(\vec{E},\vec{B})$ of electric and magnetic fields.
  This duality rotation, however, is not a symmetry of Maxwell's action.
  The first attempt to construct a duality-symmetric action was due to Zwanziger
  \cite{Zwanziger:1970hk}.
  In this construction, in order for the action to be invariant under
  the duality rotation, the manifest diffeomorphism covariance was given up.
  An alternative form of non-manifest covariance duality-symmetric action was given by Deser and Teitelboim \cite{Deser:1976iy}.
  It was later shown by \cite{Maznytsia:1998xw} that actions from \cite{Zwanziger:1970hk} and \cite{Deser:1976iy}
  are dual to each other.
  
    In the literature, the developments of 4D duality-symmetric theories
  are based mostly on actions of \cite{Zwanziger:1970hk} and \cite{Deser:1976iy}.
  For definiteness, we call duality-symmetric theories based on \cite{Deser:1976iy}
  as being in `standard formulation', whereas those based on \cite{Zwanziger:1970hk}
  are said to be in `dual formulation'.
  
  In order to
  keep the manifest diffeomorphism invariance of the action while maintaining the duality symmetry,
  an auxiliary field is introduced.
  In the standard formulation, the reference \cite{Pasti:1995tn} introduced an auxiliary scalar field.
  As for the case of dual formulation, it was studied in the reference \cite{Maznytsia:1998xw}, which introduced
  an auxiliary two-form field.
  
  In order for the field equations of
  a nonlinear extension (along with, for example, the inclusion \cite{Gaillard:1981rj}
  of axion and dilaton)
  of Maxwell's theory to be invariant under duality rotation,
  the Gaillard-Zumino condition \cite{Gibbons:1995cv,Gaillard:1997zr} has to be satisfied.
  In \cite{Carrasco:2011jv}, a procedure was constructed in order to solve Gaillard-Zumino condition,
  and hence explicitly provides examples of nonlinear theory possessing duality symmetry.
  Nonlinear duality-symmetric action in standard formulation was studied in
  \cite{Bekaert:2001wa,Bossard:2011ij,Pasti:2012wv}. In these references, a consistency condition
  relating to Gaillard-Zumino condition were derived.
  In the context of non-manifest covariant theory \cite{Bekaert:2001wa,Bossard:2011ij},
  the condition arises by demanding modified diffeomorphism invariance of the theory,
  whereas for covariant theory \cite{Pasti:2012wv},
  the condition arises by demanding the theory to be invariant
  under a local symmetry responsible for arbitrariness of auxiliary scalar field.
  
  As
  far as we are made aware, nonlinear duality-symmetric theories
  in dual formulation are less extensively studied. The reference
  \cite{Lee:2013ewa} constructed a covariantised nonlinear duality-symmetric action
  in dual formulation with the couple to axion, dilaton, and external
  electric and magnetic sources. One of our main goals is to complete such a framework.
  This includes the extension of the action of \cite{Lee:2013ewa} to describe
  a duality-symmetric action in dual formulation of a D3-brane
  coupled to 10D type IIB supergravity background.
  It could be expected that such a construction is in parallel to that
  given in the case of standard formulation.

A similar story can be seen in the case of M5-brane.
A 6D chiral two-form theory was first given
with a non-manifest covariant action \cite{Henneaux:1988gg,Perry:1996mk}.
It was then generalised and eventually reached the
complete covariant M5-brane theory \cite{Pasti:1997gx,Bandos:1997ui}.
This M5-brane theory requires an auxiliary scalar field
to maintain the worldvolume diffeomorphism invariance,
and at the same time retain non-linear self-duality.
The dual of the chiral two-form theory with quadratic action
was constructed and covariantised
by using an auxiliary four-form field \cite{Maznytsia:1998yr}.
The theory was extended to a non-manifestly covariant M5-brane theory in dual formulation \cite{Ko:2016cpw}.
It was then covariantised, by using five auxiliary scalar fields,
in the recent work \cite{Ko:2017tgo}.
The use of five auxiliary scalar fields instead of an auxiliary four-form field
makes it clear that the theory has all the desirable symmetries.
The investigation of a possibility to covariantise M5-brane in dual formulation
using other auxiliary fields still remains an open problem.

In this work, learned from the success story of the M5-brane theory in dual formulation,
we are going to construct a 4D covariant duality-symmetric theory in dual formulation
with the help of three auxiliary scalar fields, and then
demonstrate that 
interactions with other fields can be included.
Thus the theory can be improved along the same line
as its counterpart in standard formulation \cite{Pasti:2012wv}.

It will be illuminating to demonstrate the construction
on a concrete example. In particular we choose to work on
constructing in dual formulation, the covariant
duality-symmetric D3-brane worldvolume theory
in 10D type IIB supergravity background.
This will be in the Green-Schwarz formulation \cite{Green:1983wt},
in which a bosonic worldvolume is embedded
in a supergravity target space.
We will also show that,
as a standard requirement of the Green-Schwarz formulation,
the theory has kappa-symmetry which reduces the number
of fermionic degrees of freedom to match with the bosonic one.
Finally, a comparison with the standard D3-brane theory \cite{Cederwall:1996pv,Cederwall:1996ri}
will also be discussed.

This paper is organised as follows.
  In section \ref{sec:tools}, we define notation to be used in this paper
by giving basic conventions and tools.
  In section \ref{sec:covlin}, we construct covariant
quadratic 4D duality-symmetric action in dual formulation.
We also consider the inclusion of interactions with
external fields, which are coset scalars, and gravity.
The covariantisation makes use of three auxiliary scalar fields.
The theory has two local PST symmetries. One of
which can be used in order to give twisted self-duality condition.
Another one can be used to gauge-fix the auxiliary fields
and reduce the action to Zwanziger action.
  In section \ref{sec:covnonlin}, we nonlinearise the action of section \ref{sec:covlin}
and also include the interaction with external two-form fields.
Just like its counterpart in standard formulation,
in order for a nonlinear action to possess duality symmetry,
it has to satisfy a consistency condition
which is related to Gaillard-Zumino condition.
  In section \ref{Bosonsample}, we demonstrate the explicit construction
by restricting to DBI theory,
and discuss a sense in which this theory is a special case of Bossard-Nicolai theory.
  In section \ref{sec:D3}, we discuss the construction of
covariant duality-symmetric theory in dual formulation
for D3-brane theory in 10D type IIB supergravity background.
The setup from earlier sections allow this study to be carried out.
  We end the paper by giving conclusion in section \ref{sec:conclude}.

\setcounter{equation}0
\section{Basic conventions and tools}\label{sec:tools}
In this section, we define some conventions and tools
which will be used especially in sections \ref{sec:covlin}-\ref{sec:covnonlin}.
Much of the tools are followed from \cite{Pasti:2012wv}.
However, we change some conventions
to make them more suitable to this paper.
Conventions and tools other than those mentioned in this section
will also be used and will be given in later sections.

In this paper, we let the
metric signature of a 4D spacetime to be
mostly {plus} $(-,+,+,+).$
We label the coordinates as $x^\m,\ \m = 0,1,2,3,$
and let
\be
dx^\m\w dx^\n\w dx^\r\w dx^\s
=\e^{\m\n\r\s}d^4 x,
\ee
where $\e^{\m\n\r\s}$ is the Levi-Civita symbol with
$\e^{0123} = - \e_{0123} = 1.$

Consider a theory
of $N$ Abelian gauge fields $A^r(x),\ r=1,2,\cdots,N$
in 4-dimensional spacetime.
The field strengths of $A^r$ are denoted $F^r = dA^r.$
One can define the magnetic duals $A^{\rb}$
with field strengths $F^{\rb} = dA^{\rb}$
defined via
\be\label{magdefvia}
\d_{(F^r)} S = \int F^{\rb}\w \d F^r,
\ee
where $S$ is the action
constructed from the field strength $F^r,$
and
$\d_{(F^r)}$ is the variation keeping
only terms with $\d F^r.$
A $p-$form is expressed as
\be
B_{(p)} = \ove{p!}dx^{\m_1}\w\cdots \w dx^{\m_p}B_{\m_p\cdots\m_1}.
\ee
The Hodge star $\bullet$ is defined via
\be
\bullet (dx^{\m_1}\w\cdots\w dx^{\m_p})
=\frac{\sqrt{-g}}{(4-p)!}dx^{\n_4}\w\cdots\w dx^{\n_{p+1}} \e_{\r_1\cdots\r_p\n_{p+1}\cdots\n_4}g^{\m_1\r_1}\cdots g^{\m_p\r_p},
\ee
where $g^{\m\n}$ and $g$ are, respectively, inverse and determinant
of the 4D spacetime metric $g_{\m\n}.$
This gives the identity $\bullet\bullet = -1.$
With these definitions, we may write eq.\eq{magdefvia}
as
\be\label{magdef}
(\bullet F^{\bar r})^{\r\s}
=-\frac{2}{\sqrt{-g}}\frac{\d S}{\d F^r_{\r\s}}.
\ee
Let us call this equation as `duality relation'.

Next, the gauge fields $A^r$ and their duals $A^\rb$
can be collected into
\be
\vec{A}\equiv(A^i)\equiv (A^r,A^{\rb}),\qquad i=1,2,\cdots, 2N.
\ee
Likewise,
\be
\vec{F}\equiv(F^i)\equiv (F^r,F^{\rb}).
\ee
{Note that an arrow on the top of
a symbol is introduced to indicate that it is a $2N-$tuple.}
Denote $\vec{u}_i$ as a $2N-$tuple
whose $i$th entry equals to $1$
but the rest are $0.$
This allows us to express the $2N-$tuples
of gauge fields and field strengths
respectively as $\vec{A} = A^i \vec{u}_i, \vec{F}=F^i \vec{u}_i.$
For definiteness, let us denote the vector space
of $2N-$tuples as $V.$
The vector space $V$ is acted linearly by the elements
of the duality group $G\subset Sp(2N,\RR).$
The couple to gravity, scalars and fermions
can be introduced into the theory.

In particular, the scalars $\f$ parametrise a coset space $G/H$
of a principal $H-$bundle $G\to G/H,$
where $H$ is the maximal compact subgroup of $G.$
The vielbeins
on $G/H$ allows the introduction
of scalar-dependent invertible metric $M_{ij}(\f)$,
through a symmetric non-degenerate bilinear map $M$ on $V$
such that
\be
M(\vec{u}_i,\vec{u}_j) = M(\vec{u}_j, \vec{u}_i),
\ee
or $M_{ij} = M_{ji}.$
Since $G$ is a subgroup of $Sp(2N,\RR),$
one can define
an antisymmetric non-degenerate bilinear map
$\Omega$ on $V$ such that
\be
\Omega(\vec{u}_i,\vec{u}_j) = -\Omega(\vec{u}_j,\vec{u}_i),
\ee
or simply $\Omega_{ij} = -\Omega_{ji},$
and that $\Omega_{rs} = \Omega_{\bar{r}\bar{s}} = 0,\ \Omega_{r\bar{r}} = \d_{r\bar{r}}.$
One also defines $\Omega^{ij} = -\Omega^{ji}$ such that $\Omega^{ij}\Omega_{jk} = -\d^i_k.$
This allows one to define
a complex structure,
which is a linear map
$J:V\to V$ defined via
\be
\Omega(\vec{v},J\vec{w}) = -M(\vec{v},\vec{w}),\qquad
M(J\vec{v},J\vec{w}) = M(\vec{v},\vec{w}),\qquad
\Omega(J\vec{v},J\vec{w}) = \Omega(\vec{v},\vec{w}),
\ee
for all $\vec{v},\vec{w}\in V.$

As was the case of duality-symmetric theory in standard formulation
studied in \cite{Pasti:2012wv},
the study of duality-symmetric theory in dual formulation
also benefits from the use of differential form language.
As a direct extension to the $2N-$tuple of gauge fields
and field strength, one can consider the $2N-$tuple of $p-$forms.
In other words,
for $A^i_{(p)},$ a $p-$form with internal index $i,$
the associated $2N-$tuple is given by
\be
\vec{A}_{(p)} = A^{i}_{(p)}\vec{u}_i.
\ee
On these fields, define the star operation
\be
* = J\bullet,
\ee
where $J$ is a linear map on $V$
such that $J \vec{u}_i = J^j{}_i \vec{u}_j.$
So
\be
\begin{split}
*\vec{A}_{(p)}
&= J^j{}_i \bullet A_{(p)}^i \vec{u}_j,
\end{split}
\ee
or in component form
\be
(*\vec{A}_{(p)})^j = J^j{}_i \bullet A_{(p)}^i.
\ee
The star operation satisfies $** = 1.$

We define exterior derivative and interior product
to act from the right. Let $v$ be a $1-$form.
It can be shown that
\be
*i_{g^{-1}v}*\vec{A}_{(p)} = {\vec{A}_{(p)}\w v},
\ee
where $g^{-1}$ is the linear map associated
to the inverse metric of 4D spacetime.

Let us define wedge product between {two $2N-$tuple differential forms} as
\be
\vec{A}_{(p)}\w \vec{B}_{(q)}
\equiv A_{(p)}^i\w B_{(q)}^j \vec{u}_i\otimes \vec{u}_j,
\ee
where the wedge product $\w$
on RHS is the usual wedge product of differential forms.
This gives the following identities
\begin{align}\label{Omegawedgeswap}
\Omega(\vec{A}_{(p)}\w \vec{B}_{(q)})
&=-(-1)^{pq}\Omega(\vec{B}_{(q)}\w \vec{A}_{(p)}),\\
\label{Omegawedgestarflip}
\Omega(\vec{A}_{(p)}\w*\vec{B}_{(p)}) &= \Omega(\vec{B}_{(p)}\w*\vec{A}_{(p)}),\\
\Omega(\vec{A}_{(p)}\w*\vec{B}_{(p)}) &= -M(\vec{A}_{(p)}\w\bullet \vec{B}_{(p)}).
\end{align}

\setcounter{equation}0
\section{Covariant quadratic action of
4D duality-symmetric theory in dual formulation}\label{sec:covlin}
In this section, we write down and discuss the quadratic covariant action of
4D duality-symmetric theory in dual formulation.
We make use of the set up and notations given in the previous section.
In order to construct a covariant theory
in dual formulation, we make use of the idea
from \cite{Ko:2017tgo}.
This suggests the use of three auxiliary scalars\footnote{
The indices $I, J,\cdots$ are used as labels of the
auxiliary scalars. So they are clearly not
to be confused with part of the spacetime indices.
}
$a^I,\ I=0,1,2.$
Define
\be
\z^I \equiv da^I,
\ee
and the projectors
\be
P = Y^{-1}_{IJ}g^{-1}(\z^I)\otimes\z^J,\qquad
\PI = \mathbbm{1} - P,
\ee
where $Y^{-1}_{IJ}$ is the inverse of $Y^{IJ} = g^{-1}(\z^I,\z^J),$
and $\mathbbm{1}$ is the identity operator.
The projector $P$ has rank $3$ whereas the projector $\PI$
has rank $1.$
Let us denote the following induced linear transformations
\be
\cP\equiv\bigw^2 P,\qquad
\cPI\equiv\tr\lrbrk{\bigw P\bigw \PI},\qquad
\cI\equiv\bigw^2\mathbbm{1},
\ee
which transforms a tensor product of two tangent (resp. cotangent) vectors 
to a wedge product of two tangent (resp. cotangent) vectors.
For example,
\be
\cP(v_1, v_2) = Pv_1\w Pv_2,
\ee
where $v_1, v_2$ are cotangent vectors
of the 4D spacetime. Furthermore, $\tr$
means the sum of all possible permutations,
e.g.
\be
\cPI = \tr\lrbrk{\bigw P\bigw \PI}
=P\bigw\PI + \PI\bigw P,
\ee
with
\be
P\bigw\PI(v_1, v_2) = Pv_1\w \PI v_2,
\ee
and similarly for $\PI\bigw P.$
For more details, see \cite{Ko:2017tgo}.
The induced linear transformations form the following identities
\be
\cI = \cP + \cPI,\qquad
\cP\cP = \cP,\qquad
\cPI\cPI = \cPI,\qquad
\cP\cPI = \cPI\cP = 0,
\ee
\be
\cP\bigw\cPI = \cP\bigw\cI = \cI\bigw\cPI,\qquad
\cPI\bigw\cP = \cPI\bigw\cI = \cI\bigw\cP,
\ee
\be\label{sPPIs}
*\cP = \cPI*,\qquad
*\cPI = \cP*,
\ee
\be
\cP = -1 {-} \z^I\w i_{g^{-1}\z_I}.
\ee

In the study of covariant duality-symmetric theory
in dual formulation,
it will be convenient to introduce a $1-$form
\be\label{lambdef}
\l = {-}\ove{3!}\e_{I_0 I_1 I_2} \bullet(\z^{I_0}\w\z^{I_1}\w\z^{I_2}),
\ee
where
\be
\e_{I_0 I_1 I_2} = \e^{I_0 I_1 I_2} =
\begin{cases}
1 & \textrm{even permutation of }012\\
-1 & \textrm{odd permutation of }012\\
0 & \textrm{otherwise}.
\end{cases}
\ee
Let us list the following useful identities:
\be
\PI = g^{-1}(v)\otimes v,
\ee
\be
\cPI = -v\w i_{g^{-1}v},
\ee
where
\be
v = \frac{\l}{\sqrt{g^{-1}(\l,\l)}}.
\ee
Note that the equation \eq{lambdef} can also be written as
\be
\l = {-}\bullet d\lrbrk{\ove{3!}\e_{I_0 I_1 I_2} a^{I_0} \z^{I_1}\w\z^{I_2}}.
\ee
This is considered as a special case of the auxiliary field
used in \cite{Maznytsia:1998xw}, in which $\l = \bullet dB_2,$
where $B_2$ is an arbitrary $2-$form.
In \cite{Maznytsia:1998xw}, the auxiliary $2-$form $B_2$
arises from the dualisation of the auxiliary field
of the PST-covariantised duality-symmetric action in standard formulation \cite{Pasti:1995tn}.
Our choice of three $(4-1=3)$ auxiliary scalar fields $a^I$, however,
is motivated by the successful covariantisation
of M5-brane action in dual formulation \cite{Ko:2017tgo}
using five $(6-1 = 5)$ auxiliary scalar fields,
which in turn is a specific restriction, motivated by a basic geometrical argument,
of the auxiliary $4-$form.

In the subsequent calculations, we will
also need to make use of the following identities for variations with respect to $a^I.$
By using the result
\be\label{dPda}
\d_{(a)}P = \PI g^{-1}\d\z^I \otimes\z_I + g^{-1}\z_I \otimes\PI\d\z^I,
\ee
one obtains
\be\label{dcPda}
\begin{split}
\d_{(a)}\cP \vec{F}
&=*(\PI\d\z^I\w i_{g^{-1}\z_I}*\vec{F}) {-} \PI\d\z^I\w i_{g^{-1}\z_I}\vec{F},
\end{split}
\ee
where
\be
\z_I \equiv Y^{-1}_{IJ}\z^J.
\ee
Let us also note another useful identity:
\be\label{cLcPcF}
\cL_{g^{-1}\z_I}(\cP\vec{\cF})\w\cP\vec{\cF} = 0.
\ee

The covariant duality-symmetric quadratic action in dual formulation is given by
\be\label{quadS}
S = -\ove{2}\int\Omega(\vec{F}\w\cP\vec{\cF}),
\ee
or equivalently,
\be
\begin{split}
S
&= -\ove{4}\int M(\vec F\w\bullet \vec F) + \ove{4}\int M(i_{g^{-1}v}\vec\cF \w \bullet i_{g^{-1}v}\vec\cF),\\
&= -\ove{8}\int d^4x\sqrt{-g} M_{ij}(\f) F^i_{\m\n}F^{j\m\n} - \ove{4}\int d^4 x\sqrt{-g} M_{ij}(\f)\cF^i_{\m\n}v^\n \cF^{j\m\s} v_\s,
\end{split}
\ee
where $\vec\cF = \vec F-*\vec F,$
and the hatted Roman indices are for spatial part of the 4D spacetime:
$\ah,\bh,\cdots = 1,2,3.$
The form of the action makes it clear that
we allow the gauge fields to interact
with coset scalars and 4D gravity.
However, we do not consider dynamics of scalars and gravity,
i.e. we treat them as external fields.
In the next sections, we will also allow
the interaction with other external fields.

Using the identities \eq{Omegawedgeswap}, \eq{Omegawedgestarflip},
and \eq{sPPIs}, one obtains the variation of the action with respect to $A^i:$
\be\label{dSdAquad}
\d_{(\vec{A})}S = \int\Omega(\d \vec A\w d\cP\vec\cF) - \ove{2}\int d(\Omega(\d \vec A\w(2\cP\vec\cF-\vec F))).
\ee
To compute the variation of the action with respect to $a^I,$
we use the identities \eq{dPda}, \eq{dcPda},
and \eq{cLcPcF}, which give
\be\label{dSdaquad}
\begin{split}
\d_{(a)}S
&= -\int\Omega\lrbrk{\d a^I i_{g^{-1}\z_I}\cP\vec\cF\w d\cP\vec\cF}
+\ove{2}\int d\lrbrk{\Omega(\d a^I i_{g^{-1}\z_I}\cP\vec\cF\w\cP\vec\cF)}.
\end{split}
\ee

Using the above variations, one obtains
the field equations of $A^i$
\be\label{EOMAquad}
d\cP \cF^i = 0.
\ee
One also obtains the field equations for $a^I$
\be\label{EOMaquad}
\Omega_{ij}i_{g^{-1}\z_I}\cP\cF^i\w d\cP\cF^j= 0,
\ee
which are trivially satisfied after imposing
field equations \eq{EOMAquad} for $A^i.$
This suggests that the auxiliary fields
have no dynamics of their own,
reflecting their auxiliary nature.

The quadratic action \eq{quadS} possesses two important
types of local symmetries. These are called PST1 and PST2.
The PST1 symmetry is given by
\be\label{PST1quad}
\d A^i = \psi_I^i(a^K)da^I,\qquad
\d a^I = 0,
\ee
where $\psi_I^i$ are arbitrary functions of
auxiliary fields $a^I.$ The form \eq{PST1quad} suggests
that PST1 is semi-local \cite{Bekaert:1998yp,Bandos:2014bva}.
In the case where it is a gauge symmetry,
it can be used to gauge-fix the field equations to give
the twisted self-duality condition.
The PST2 symmetry is given by
\be\label{PST2quad}
\d a^I = \vphi^I,\qquad
\d \vec A = \vphi^I i_{g^{-1}\z_I}\cP\vec\cF.
\ee
It is used to ensure that the auxiliary fields $a^I$
are arbitrary.

To analyse the semi-local PST1 symmetry, let us follow
the analysis of \cite{Bandos:2014bva}.
A criteria to determine whether a symmetry is gauge
or global symmetry is to check the vanishing
of Noether's charge. If the Noether's charge vanishes,
then the symmetry is a gauge symmetry (see for example \cite{Henneaux:1992ig}).
The Noether's current of PST1 symmetry is given by
\be\label{Noethercurrent}
\begin{split}
j
&= \bullet\Omega(\vec\psi_I da^I\w\cP\vec\cF).
\end{split}
\ee
One then investigates the dynamical system of action \eq{quadS}
to look for the branch in which the Noether's charge vanishes.
Since the action is singular when $g^{-1}(\l,\l) = 0,$
it turns out that the dynamical system is separated into two branches.
One of them has $g^{-1}(\l,\l) > 0,$ while the other has
$g^{-1}(\l,\l) < 0.$ These two branches are disconnected
because there is no smooth PST2 transformation
which joins the two branches without passing through the forbidden
$g^{-1}(\l,\l) = 0$ region.
The Noether's charge of PST1 vanishes if $\l^0 = 0.$
This is the case only in the $g^{-1}(\l,\l) > 0$ branch,
in which PST2 symmetry can be used to gauge-fix the auxiliary fields as, for example,
$a^I = x^I.$ Therefore, in this branch, the PST1 symmetry
is a gauge symmetry. So the field equation is equivalent
to the twisted self-duality condition
\be
F^i = *F^i.
\ee
On the other hand, in the $g^{-1}(\l,\l) < 0$ branch,
the Noether's charge is always zero. So in this branch
one does not obtain the twisted self-duality condition.

In the $g^{-1}(\l,\l) > 0$ branch, which is permissible,
one can use PST2 symmetry \eq{PST2quad}
to fix the gauge
\be\label{gaugefixa}
a^I = (\z_0)_\m^I x^\m,
\ee
where $(\z_0)_\m^I$ are constants, such that after substituting
eq.\eq{gaugefixa} into eq.\eq{lambdef},
one obtains spacelike $\l.$
The action becomes
\be\label{Zwangeneral}
S
=-\ove{8}\int d^4x\sqrt{-g} M_{ij}(\f) F^i_{\m\n}F^{j\m\n} - \ove{4 g^{-1}(n,n)}\int d^4 x\sqrt{-g} M_{ij}(\f)\cF^i_{\m\n}n^\n \cF^{j\m\s} n_\s,
\ee
where
\be
n^\n = (\z_0)^0_{\m_0}(\z_0)^1_{\m_1}(\z_0)^2_{\m_2}\frac{\e^{\n\m_0\m_1\m_2}}{\sqrt{-g}}.
\ee
From the action \eq{Zwangeneral}, if one turns off coset scalars, and gravity,
then one obtains a source-free version of Zwanziger's action \cite{Zwanziger:1970hk}.
In the gauge \eq{gaugefixa},
under the combined local transformation of PST2
and 4D diffeomorphism $\d x^\m = \xi^\m(x)$
the auxiliary fields transform as
\be
\d a^I(x) = \xi^\m(x)\pa_\m a^I(x) + \vphi^I(x) = \xi^\m(x)(\z_0)^I_\m + \vphi^I(x).
\ee
In order for this combined transformation
to not modify the gauge-fixing condition \eq{gaugefixa},
the PST2 gauge parameter has to be chosen as
\be
\vphi^{I}(x) = -\xi^\m(x)(\z_0)^I_\m,
\ee
in which case, the combined local transformation
on $A_{\m}^i$
is given by
\be\label{moddiffeo}
\d A_{\m}^j =
\xi^\r\pa_\r A_{\m}^j + \pa_{\m} \xi^\r A_{\r}^j 
- \ove{g^{-1}(n,n)}\xi^\s J^j{}_i\cF^{i\r}{}_\l n^\l n^\n \e_{\m \n\r\s}\sqrt{-g},
\ee
which is the modified diffeomorphism of the action \eq{Zwangeneral}.
Note that this transformation does not depend explicitly
on the choice of $(\z_0)^I_\m.$

\setcounter{equation}0
\section{Covariant nonlinear action of
4D duality-symmetric theory in dual formulation}\label{sec:covnonlin}
In this section, we nonlinearise the action \eq{quadS}.
Furthermore, we also include interactions
with other fields, particularly two-form fields $\vec{C}_2.$
We start by writing down the action
\be\label{covnonlingen}
S_{cds} = S_1 + S_2 +S_3,
\ee
where
\be
S_1 = -\ove{2}\int\Omega(\vec F\w\cP \vec F),\qquad
S_2 = \int d^4 x\sqrt{-g}\cL,\qquad
S_3 = \ove{2}\int \Omega(\vec F\w \vec C_2),
\ee
where now $F^i = dA^i - C_2^i,$
and we consider the case in which
$A^i$ and $a^I$ appear in $\cL$
only through $\cPI F^i.$
The variation of the action with respect to $\vec A$ is given by
\be
\d_{(\vec A)}S_{cds} = \d_{(\vec A)}S_1 + \d_{(\vec A)}S_2 +\d_{(\vec A)}S_3,
\ee
where
\be\label{dSdAnonl1}
\d_{(\vec A)}S_1 = \int\Omega(\d \vec A\w d\cP \vec F)
+\ove{2}\int\Omega(\d \vec A\w d\vec C_2)
 - \ove{2}\int d(\Omega(\d \vec A\w(2\cP \vec F-\vec F))),
\ee
\be\label{dSdAnonl2}
\d_{(\vec A)}S_2 = -\int\Omega(\d \vec A\w d\cP*\vec X) + \int d(\Omega(\d \vec A\w\cP*\vec X)),
\ee
\be\label{dSdAnonl3}
\d_{(\vec A)}S_3 = -\ove{2}\int\Omega(\d \vec A\w d\vec C_2) + \ove{2}\int d(\Omega(\d \vec A\w \vec C_2)),
\ee
where
\be
X^i = \hlf dx^\m\w dx^\n X^i_{\n\m},
\ee
\be
X^{i\m\n} = -2M^{ij}\frac{\d\cL}{\d(\cPI F)^j_{\m\n}}.
\ee
Note that $\cP \vec X = 0.$
The variation with respect to $a^I$
can be conveniently obtained by using an identity
\be
\Omega(d(\cP \vec F_1\w i_{g^{-1}\z_I}\cP \vec F_2))
={-}\Omega(d\cP \vec F_1 \w i_{g^{-1}\z_I}\cP \vec F_2 + d\cP \vec F_2 \w i_{g^{-1}\z_I}\cP \vec F_1),
\ee
giving
\be\label{dSdanonl}
\begin{split}
\d_{(a)}S_{cds}
&=\ove{2}\int \d a^I d\lrbrk{\Omega\lrbrk{\cP* \vec F\w i_{g^{-1}\z_I}\cP*\vec F - \cP* \vec X\w i_{g^{-1}\z_I}\cP*\vec X}}\\
&\quad-\int\d a^I \Omega\lrbrk{d\cP(\vec F-*\vec X)\w i_{g^{-1}\z_I}\cP(\vec F-*\vec X)}\\
&\quad-\ove{2}\int d\lrbrk{\d a^I\Omega\lrbrk{\cP*\vec F\w i_{g^{-1}\z_I}\cP*\vec F + \cP \vec F\w i_{g^{-1}\z_I}\cP \vec F
-2\cP \vec F\w i_{g^{-1}\z_I}\cP*\vec X}}.
\end{split}
\ee
Note that for the quadratic action \eq{quadS},
the first term on the RHS of eq.\eq{dSdanonl}
vanishes. So for the nonlinear case, let us demand
\be\label{dcGZ}
d\lrbrk{\Omega\lrbrk{\cP* \vec F\w i_{g^{-1}\z_I}\cP*\vec F - \cP* \vec X\w i_{g^{-1}\z_I}\cP*\vec X}}
=0.
\ee
This condition will ensure the existence of PST2 symmetry.
It is also analogous to the condition obtained in \cite{Pasti:2012wv},
which in turn is the covariantisation of the conditions by \cite{Bekaert:2001wa,Bossard:2011ij},
and is also related to the Gaillard-Zumino condition
\cite{Gibbons:1995cv,Gaillard:1997zr},
which requires the consistency of duality transformation \cite{Gaillard:1981rj}
on duality-symmetric theory.

The field equations for $\vec A, a^I,$
can be read off from the variations \eq{dSdAnonl1}, \eq{dSdAnonl2}, \eq{dSdAnonl3}, \eq{dSdanonl},
which respectively give
\be
d\cP(\vec F-*\vec X) = 0,
\ee
\be
\Omega_{ij}i_{g^{-1}\z_I}\cP(\vec F-*\vec X)^i\w d\cP(\vec F-*\vec X)^j
= 0.
\ee
Clearly, the field equation for $a^I$
is implied by the field equation for $\vec A.$
The variations \eq{dSdAnonl1}, \eq{dSdAnonl2}, \eq{dSdAnonl3}, \eq{dSdanonl}
can also be used to determine the PST1
and PST2 symmetries.
PST1 symmetry is the same
as that, eq.\eq{PST1quad}, of quadratic action:
\be
\d A^i = \psi_I^i(a^K)da^I,\qquad
\d a^I = 0,\qquad
\ee
where $\psi_I^i$ is an arbitrary function of
auxiliary fields $a^I.$
In case where PST1 is a gauge symmetry,
it allows gauge fixing of $A^i$ field equations
giving non-linear twisted self-duality condition
\be
\vec{X} = *\vec{F}.
\ee
The PST2 symmetry is given by
\be
\d a^I = \vphi^I,\qquad
\d \vec A = \vphi^I i_{g^{-1}\z_I}\cP(\vec F-*\vec X).
\ee
Its presence ensures the arbitrariness of the auxiliary fields $a^I.$

In
order to determine the case in which PST1 is a gauge symmetry,
we can again follow \cite{Bandos:2014bva}.
By the analysis similar to that given in section \ref{sec:covlin},
one can also conclude that PST1 is a gauge symmetry
only in the $g^{-1}(\l,\l)>0$ branch.
However, in the $g^{-1}(\l,\l)<0$ branch, which is separated from the $g^{-1}(\l,\l)>0$ branch,
one does not obtain non-linear twisted self-duality condition.
Note that this analysis will also hold in later sections of this paper.

Let us now proceed to simplify the condition \eq{dcGZ}.
For any $2N-$tuples of $1-$forms $\vec A, \vec B,$
we define the following symmetric bilinear bracket
\be
(\vec A,\vec B)^\s = \ove{\sqrt{-g}}\Omega(\vec A_\m \vec B_\n v_\r \e^{\m\n\r\s}) = (\vec B,\vec A)^\s.
\ee
Let $K^\m_\n$ be symmetric and satisfies $K^\m_\n v_\m = 0.$
We have the following identity
\be\label{AKB}
(\vec A,K\vec B)^\s + (\vec B,K\vec A)^\s = -(\vec A,\vec B)^\n(\Delta K)^\s_\n,
\ee
where $K\vec A = K^\m_\n A^i_\m dx^\n \vec u_i,$
and $(\Delta K)^\m_\n = K^\m_\n - [K]\d^\m_\n,$
with square bracket $[\cdot]$ denoting the
trace over spacetime indices.
Direct computation gives
\be
\bullet\Omega(\cP*\vec X\w i_{g^{-1}\z_I}\cP*\vec X)
=-(i_{g^{-1}v}\vec X,i_{g^{-1}v}\vec X)^\b\z_{I\b}v_\m dx^\m,
\ee
and
\be
{\bullet\Omega(\cP*\vec F\w i_{g^{-1}\z_I}\cP *\vec F)}
=-(\vec f,\vec f)^\b\z_{I\b}v_\m dx^\m,
\ee
where $\vec f = i_{g^{-1}v}(\cPI \vec F).$
One can see that the condition \eq{dcGZ} is satisfied
if
\be\label{dcGZreduced}
(i_{g^{-1}v}\vec X,i_{g^{-1}v}\vec X)^\b = (\vec f,\vec f)^\b.
\ee

\setcounter{equation}0
\section{DBI action in covariant duality-symmetric theory in dual formulation}\label{Bosonsample}

In this and the next section, we will discuss explicit examples
by restricting to $N=1,$ i.e. duality-symmetric actions
will contain one pair of gauge fields.
In particular, we will discuss DBI, and also D3-brane.
For these examples, we will relate the one-potential
actions (i.e. those containing only `electric' gauge field
but not `magnetic' one) with
the covariant duality-symmetric actions in dual formulation.
We will relate the duality relation \eq{magdef}
with non-linear twisted self-duality condition.
We will also compare the on-shell actions.

By setting $N=1,$ the internal indices $i,j,k,\cdots$
take values only in $\{1,2\}.$
In this case, there is the following decomposition
\be\label{CH}
(\vec f^{\,5})^i_\m = [\vec f^{\,2}](\vec f^{\,3})^i_\m - \hlf\lrbrk{[\vec f^{\,2}]^2 - [\vec f^{\,4}]}f^i_\m,
\ee
where
\be
(\vec f^{\,2})^\m_\n = f^{i\m}M_{ij}f^j_\n,\qquad
(\vec f^{\,3})_\m^i = (\vec f^{\,2})^\n_\m f_\n^i,\qquad
(\vec f^{\,4})^\m_\n = (\vec f^{\,2})^\m_\r(\vec f^{\,2})^\r_\n,\quad etc.
\ee
Note also that by using the identity
\be
\Omega^{ij}M_{jk}\Omega^{kl}M_{lm} = -\d^i_m,
\ee
one obtains $\det(M_{ij}) = 1.$

We are interested in $\cL$ which depends on
$(\cPI \vec F)^i_{\m\n}$ but not on
any derivatives of $(\cPI \vec F)^i_{\m\n}.$
Based on the equation \eq{CH}, there are only two invariants
and are given by
\be
U\equiv -\ove{4}(\cPI \vec F)^i_{\m\n}(\cPI \vec F)^{j\m\n}M_{ij}
=-\hlf[\vec f^{\,2}],
\ee
\be
Y\equiv \ove{32}M_{ij}(\cPI \vec F)^i_{\m\n}(\cPI \vec F)^j_{\r\s}
M_{kl}(\cPI \vec F)^{k\m\n}(\cPI \vec F)^{l\r\s}
=\ove{8}[\vec f^{\,4}].
\ee
The next step is to write $\cL$ as a polynomial of $U$ and $Y,$
and note the identity
\be
i_{g^{-1}v}\vec X = \vec\cD\cL,
\ee
where differential operator $\vec\cD$ is defined as
\be
\vec\cD \cdot = 2 M^{ij}\frac{\d \cdot}{\d(\cPI \vec F)^{j\m\n}}v^\n dx^\m \vec u_i.
\ee
Applying $\vec\cD$ to the invariants gives
\be
\vec\cD U = \vec f,
\qquad
\vec\cD Y = -\hlf\vec f^{\,3},
\ee
where
\be
\vec f^{\,3} = (\vec{f}^{\,3})^i_\m dx^\m \vec u_i.
\ee
Therefore,
\be
\begin{split}
i_{g^{-1}v}\vec X
&= \frac{\d\cL}{\d U} \vec f -\hlf \frac{\d\cL}{\d Y} \vec f^{\,3}.
\end{split}
\ee
By using the identities
\be
(\vec f,\vec f^{\,3})^\s = \hlf [\vec{f}^{\,2}](\vec f,\vec f)^\s,
\ee
\be
(\vec f^{\,3},\vec f^{\,3})^\s = \hlf\lrbrk{[\vec f^{\,2}]^2 - [\vec f^{\,4}]}(\vec f,\vec f)^\s,
\ee
which are obtained from eq.\eq{AKB},
we obtain
\be
(i_{g^{-1}v}\vec X,i_{g^{-1}v}\vec X)^\s
=\lrsbrk{\lrbrk{\frac{\d\cL}{\d U}}^2 + \frac{\d\cL}{\d U}\frac{\d\cL}{\d Y} U + \lrbrk{\frac{\d\cL}{\d Y}}^2\lrbrk{\hlf U^2-Y}}(\vec f,\vec f)^\s.
\ee
The condition \eq{dcGZreduced} demands
\be\label{PSlike}
\lrbrk{\frac{\d\cL}{\d U}}^2 + \frac{\d\cL}{\d U}\frac{\d\cL}{\d Y} U + \lrbrk{\frac{\d\cL}{\d Y}}^2
\lrbrk{\hlf U^2-Y}
=1.
\ee
The equation of this and related form
also appears in several papers on 4D nonlinear
duality-symmetric theory \cite{Deser:1997gq,Bekaert:2001wa,Bossard:2011ij,Pasti:2012wv},
as well as the ones on M5-brane \cite{Perry:1996mk,Schwarz:1997mc,Bekaert:1998yp}.
See also \cite{Ko:2013dka}
for analogous equations for M5-brane.
It is a crucial requirement for the nonlinear deformation
of non-manifest covariant 4D duality-symmetric theory and non-manifest covariant 6D chiral 2-form theory
to be diffeomorphism invariant.
Or for the PST-covariantised counterparts, the requirement arises
by demanding that nonlinear deformation has PST2 symmetry.

The equation \eq{PSlike} takes exactly the same form
as the one analysed in \cite{Perry:1996mk}
in the context of M5-brane theory.
In fact, it was this form of the equation that
was crucial to achieve the DBI-like form of M5-brane action.
The analysis gave two special solutions
which are $\cL = U,$ which in our case corresponds to the quadratic action
of duality-symmetric theory in dual formulation.
The second special solution reads, after some trivial
redefinitions,
\be\label{cLDBIlike}
\cL = \frac{1}{\b}\lrbrk{1-\sqrt{1-2\b U + 4\b^2\lrbrk{\hlf U^2 - Y}}}.
\ee
Substituting this into eq.\eq{covnonlingen}
gives a DBI-like nonlinear deformation of 
duality-symmetric action in dual formulation
\be\label{cdsDBI}
\begin{split}
S_{cds-DBI}
&= -\ove{2}\int\Omega(\vec F\w\cP \vec F)
+\int d^4 x\frac{\sqrt{-g}}{\b}\lrbrk{1-\sqrt{1-2\b U + 4\b^2\lrbrk{\hlf U^2 - Y}}}\\
&\qquad+\hlf\int\Omega(\vec{F}\w\vec{C}_2)
\end{split}
\ee
For this theory, the nonlinear self-duality equation reads
\be\label{DBIlikeEOM}
\vec f_*\equiv i_{g^{-1}v}\cPI*\vec F = \frac{(1-2\b U)\vec f-\b \vec f^{\,3}}{\sqrt{1-2\b U + 4\b^2\lrbrk{\hlf U^2 - Y}}}.
\ee
Note
that in the case where the 4D metric is flat and $\vec{C}_2 = 0$, and with identification
\be
(M_{ij})
=
\begin{pmatrix}
e^{-\f} + a^2 e^\f & -ae^\f\\
-ae^\f & e^{\f}
\end{pmatrix}
,
\ee
the action \eq{cdsDBI} reduces to the source-free version of the action given in \cite{Lee:2013ewa}.
It will be possible to see in the next section how
this form of the action is extended to the complete description of covariantised
duality-symmetric D3-brane action in dual formulation coupled to 10D type IIB supergravity.

There are also other possible solutions to eq.\eq{PSlike}.
Bossard and Nicolai \cite{Bossard:2011ij} present series solutions to this equation.
Let us give a brief review
using our notation.
Defining $Z =4Y-U^2$ the equation \eq{PSlike} becomes
\be\label{PSlikeBN}
\lrbrk{\frac{\d\cL}{\d U}}^2 - 4\lrbrk{\frac{\d\cL}{\d Z}}^2 Z = 1.
\ee
After imposing an ansatz, dubbed BN ansatz,
\be\label{BNansatz}
\cL = U + \sum_{n=0}^\infty\ove{(2+2n)!}H^{(n)}(U)Z^{1+n},
\ee
the condition \eq{PSlikeBN} will give
the following conditions
\be\label{BNcondfromZ1}
\frac{\pa H^{(0)}}{\pa U} = ( H^{(0)})^2,
\ee
\be\label{BNcondfromZ2}
\begin{split}
&2(H^{(0)})^{2+2k}\frac{\pa }{\pa U}\lrbrk{\frac{H^{(k)}}{(H^{(0)})^{2+2k}}}
\\
&=\lrbrk{\sum_{n=1}^{k-1}\binom{2+2k}{1+2n}H^{(n)}H^{(k-n)}
-\sum_{n=0}^{k-1}\binom{2+2k}{2+2n}\frac{\pa H^{(n)}}{\pa U}\frac{\pa H^{(k-n-1)}}{\pa U}};\ k=1,2,3,\cdots.
\end{split}
\ee
This allows the recursive procedure leading to
the solution of the form
\be\label{BNsoln}
H^{(k)} = c^{(k)}(H^{(0)})^{2+2k}+\Ht^{(k)},
\ee
where $\Ht^{(k)}$ is a particular solution,
and $H^{(0)}$ is a solution to eq.\eq{BNcondfromZ1}.
By following the iterative procedure,
the reference \cite{Bossard:2011ij} demonstrated
how to obtain a non-linear deformation
of 4D duality-symmetric theory.
The action obtained this way does not necessarily
take the DBI-like form. In fact, it is not known
whether there are other actions having closed form, apart from the DBI-like form.

Although the analysis was originally given in the standard formulation,
the above discussion shows, by closely following the analysis in \cite{Bossard:2011ij},
explicitly that the analysis for the covariant duality-symmetric theory
in the dual formulation can be done along the very similar line.
This is thanks to the equation of the form \eq{PSlike},
which are shared in both formulations
(but with $U$ and $Y$ defined differently in different formulations).

The BN ansatz is more general than the DBI-like form eq.\eq{cLDBIlike}
in the sense that the latter can be obtained as a special case of the former.
In order to see this in more details,
let us start by rewriting
the equation \eq{cLDBIlike} in terms of the invariants $U,Z.$
So
\be
\cL = \frac{1}{\b}\lrbrk{1-\sqrt{\lrbrk{1-\b U}^2 - \b^2 Z}}.
\ee
Expanding in $Z$ gives
\be
\begin{split}
\cL
&= U + \sum_{n=0}^\infty\frac{(2n)!}{2^{2n+1} n!(n+1)!}\lrbrk{\frac{\b}{1-\b U}}^{2n+1} Z^{1+n}.
\end{split}
\ee
This allows us to read off
\be
\begin{split}
H^{(n)}
&= (2n-1)!!(2n+1)!!(H^{(0)})^{2n+1},
\end{split}
\ee
where
\be
H^{(0)} = \frac{\b}{1-\b U}.
\ee
Comparing with the form \eq{BNsoln},
it can be concluded that
the DBI-like nonlinearisation
is a special case of BN nonliearisation
with $c^{(k)} = 0$ for all $k=1,2,3,\cdots.$
From this observation, we see that the DBI-like nonlinear deformation
is the most special among the nonlinear deformations
in the sense that it is obtained from the BN ansatz
in which the solutions $H^{(k)}, k=1,2,3,\cdots$
to the differential equations \eq{BNcondfromZ2} do not have
non-trivial complementary solution.

Duality-symmetry does not completely fix
the nonlinear deformation.
This is also the case for duality-symmetric theories with $\cN=1,2$
rigid supersymmetry \cite{Carrasco:2011jv,Broedel:2012gf}.
The main goal of our paper, however, is not to pursue
nonlinearly deformed action other than the standard DBI-like one.
Let us therefore focus only on DBI-like actions from now on.

As to be expected from the literature, see for example \cite{Deser:1997gq},
it should hold that the DBI-like duality-symmetric equation
\eq{DBIlikeEOM} coincide with the standard DBI equation.
Let us proceed to show this.

Using the identity
\be
\lrbrk{\mathbbm{1}+\b \vec f^{\,2}}\lrbrk{(1-2\b U)\mathbbm{1}-\b \vec f^{\,2}}\vec{f}
=\lrbrk{1-2\b U + 4\b^2\lrbrk{\hlf U^2 - Y}}\vec{f},
\ee
we obtain
\be\label{DBIlikef}
\begin{split}
f^i_\m
&= \frac{(f_*)^i_\m+\b (\vec f^{\,2})^i_j (f_*)^j_\m}{\sqrt{1-2\b U + 4\b^2\lrbrk{\hlf U^2 - Y}}}.
\end{split}
\ee
Let us consider the special case where $\vec C_2 = \vec{0},$ no $G/H-$coset scalar,
and that $(M_{ij}) = \diag(1,1).$
A more general example will be discussed in the next section.
Let us denote $F^1\equiv F, F^2\equiv G, f^1\equiv f, f^2\equiv g,$
i.e. ${f_\m = v^\n F_{\n\m}, g_\m = v^\n G_{\n\m}.}$
We also define
\be
\Ft_{\m\n} \equiv \hlf\sqrt{-g}\e_{\m\n\r\s}F^{\r\s},\qquad
\Gt_{\m\n} \equiv \hlf\sqrt{-g}\e_{\m\n\r\s}G^{\r\s},
\ee
and ${\ft_\m \equiv v^\n\Ft_{\n\m}, \gt_\m \equiv v^\n\Gt_{\n\m}.}$
Note that
\be
\bullet F = -\hlf \Ft_{\m\n} dx^{\n\m}\equiv - \Ft,
\ee
and similarly, $\bullet G = -\Gt.$
Let us also note the following useful identities
\be\label{idenqtqt}
\ft^\m \ft_\m = \hlf [F^2] + f_\m f^\m,
\ee
\be\label{idenqqt}
f^\m \ft_\m = -\ove{4} [F\Ft].
\ee
With the above set up, we can rewrite \eq{DBIlikeEOM}
and \eq{DBIlikef}
in matrix notation as
\be\label{DBIlikeft2}
-\gt = \frac{(1-2\b U)f - \b([f^2]f + [fg]g)}{\sqrt{1-2\b U + 4\b^2\lrbrk{\hlf U^2 - Y}}},\qquad
\ft = \frac{(1-2\b U)g - \b([fg]f + [g^2]g)}{\sqrt{1-2\b U + 4\b^2\lrbrk{\hlf U^2 - Y}}},
\ee
\be\label{DBIlikef2}
f = \frac{-\gt + \b(-[f^2]\gt + [fg]\ft)}{\sqrt{1-2\b U + 4\b^2\lrbrk{\hlf U^2 - Y}}},\qquad
g = \frac{\ft + \b(-[fg]\gt + [g^2]\ft)}{\sqrt{1-2\b U + 4\b^2\lrbrk{\hlf U^2 - Y}}},
\ee
where $U$ and $Y$ can be expressed as
\be
U = -\hlf [f^2] - \hlf [g^2],\qquad
Y = \ove{8}[f^2]^2 + \ove{4}[fg]^2 + \ove{8}[g^2]^2.
\ee
The strategy to solve the equations \eq{DBIlikeft2}-\eq{DBIlikef2}
is to first solve it perturbatively in $\b$ and look for the pattern
of the solution which allows us to guess
the closed form, and then check if the guessed closed form
is indeed the solution.
Let us start the process by expanding the equations \eq{DBIlikeft2}-\eq{DBIlikef2}
in $\b$ and iteratively solving order-by-order,
trading $[fg], [g^2]$ for $[F^2], [F\Ft].$
As a concrete demonstration, we present the result at order $\b^2:$
\be
-\gt = f + \frac{\b}{4}\lrbrk{[F^2]f + [F\Ft]\ft}
+ \frac{\b^2}{32}\lrbrk{(3[F^2]^2 + [F\Ft]^2)f + 2[F^2][F\Ft]\ft}+\cO(\b^3),
\ee
\be
g = \ft + \frac{\b}{4}\lrbrk{[F^2]\ft - [F\Ft]f}
+ \frac{\b^2}{32}\lrbrk{(3[F^2]^2 + [F\Ft]^2)\ft - 2[F^2][F\Ft]f}+\cO(\b^3).
\ee
Next, we use the identities $F = {-}v\w f {-} \bullet(v\w\ft), G = {-}v\w g {-} \bullet(v\w\gt)$
to obtain
\be
-\Gt = F + \frac{\b}{4}\lrbrk{[F^2]F + [F\Ft]\Ft}
+ \frac{\b^2}{32}\lrbrk{(3[F^2]^2 + [F\Ft]^2)F + 2[F^2][F\Ft]\Ft}+\cO(\b^3).
\ee
By making the perturbative calculation up to order, say $\b^{9},$
it is evident that
\be
\Gt = -\frac{F + \frac{\b}{4}\Ft [F\Ft]}{\sqrt{1-\frac{\b}{2}[F^2]-\frac{\b^2}{16}[F\Ft]^2}}+\cO(\b^{10}),
\ee
which up to this order, is a DBI duality condition.

Having obtained the guessed form of the duality condition,
let us now outline how to show that the equations \eq{DBIlikeft2}-\eq{DBIlikef2}
indeed agree with
\be\label{DBIdualcond}
\Gt = -\frac{F + \frac{\b}{4}\Ft [F\Ft]}{\sqrt{1-\frac{\b}{2}[F^2]-\frac{\b^2}{16}[F\Ft]^2}}.
\ee
We start by substituting this equation
into the expression of $U,Y$
giving
\be
U = -\frac{[f^2]+\ove{4}[F^2]-\frac{\b}{4}[f^2][F^2]+\frac{\b}{16}[F\Ft]^2}{1-\frac{\b}{2}[F^2]-\frac{\b^2}{16}[F\Ft]^2},
\ee
\be
\hlf U^2 - Y =
-\frac{-\ove{4}[f^2]^2-\ove{8}[f^2][F^2]+\frac{1}{64}[F\Ft]^2}{1-\frac{\b}{2}[F^2]-\frac{\b^2}{16}[F\Ft]^2}.
\ee
This in turn gives
\be
1-2\b U + 4\b^2\lrbrk{\hlf U^2 - Y}
=\frac{\lrbrk{1+\b[f^2]}^2}{1-\frac{\b}{2}[F^2]-\frac{\b^2}{16}[F\Ft]^2}.
\ee
It is also useful to compute
\be
[fg] = -\frac{\lrbrk{\ove{4}+\frac{\b}{4}[f^2]}[F\Ft]}{\sqrt{1-\frac{\b}{2}[F^2]-\frac{\b^2}{16}[F\Ft]^2}},
\qquad
[g^2] = \frac{[f^2] + \hlf[F^2] + \frac{\b}{8}[F\Ft]^2 + \frac{\b^2}{16}[f^2][F\Ft]^2}{{1-\frac{\b}{2}[F^2]-\frac{\b^2}{16}[F\Ft]^2}}.
\ee
Using these results, eq.\eq{DBIlikeft2}-\eq{DBIlikef2}
can easily be shown to hold.

We have thus shown that the DBI-like duality-symmetric field equation \eq{DBIlikeEOM}
can be put in the form
\be
\Gt = -\frac{F + \frac{\b}{4}\Ft [F\Ft]}{\sqrt{1-\frac{\b}{2}[F^2]-\frac{\b^2}{16}[F\Ft]^2}},
\ee
which is the duality relation corresponding to the action
\be
S_{DBI} = \int d^4 x\ove{\b}\lrbrk{\sqrt{-g}-\sqrt{-\det(g_{\m\n} + \sqrt{\b}F_{\m\n})}},
\ee
i.e.
\be
\Gt^{\m\n} = \frac{2}{\sqrt{-g}}\frac{\d S_{DBI}}{\d F_{\m\n}},
\ee
which is of the form \eq{magdef}.
Furthermore, it is interesting to compare the on-shell action.
By `on-shell', we impose the condition \eq{DBIdualcond}
which was shown earlier to be equivalent to non-linear twisted self-duality condition.
By a simple substitution, we find that
\be\label{cdsDBIonshell}
S_{cds}
\stackrel{\text{on-shell}}{=}
S_{DBI} - \ove{2}\int F\w G.
\ee
In fact, if we also impose DBI field equation $\pa_\m(\sqrt{-g}\Gt^{\m\n}) = 0,$
then the second term on the RHS of eq.\eq{cdsDBIonshell} is simply a surface term.
The on-shell agreement of the DBI-like duality-symmetric action in the standard formulation
and the DBI action was done in \cite{Berman:1997iz}.
We have also shown that, as expected, the agreement
also holds in the case of dual formulation of duality-symmetric theory.

Having obtained the closed form of the action of the duality-symmetric theory
corresponding to DBI action, it is natural to expect that
the duality-symmetric theory corresponding to supersymmetric D3-brane
in Green-Schwarz formalism can also be obtained.
We will show in the next section how this is constructed.

\setcounter{equation}0
\section{Covariant duality-symmetric theory in dual formulation of kappa-symmetric D3-brane}\label{sec:D3}
In Green-Schwarz formalism, a D3-brane worldvolume is embedded
in a target superspace given by type IIB supergravity.
The gauge field $A$ lives on the D3-brane worldvolume.
It couples to the pullbacks of the following superfields on target superspace.
These fields are supervielbeins, a dilaton, an NS-NS $B$-field,
and RR C-fields.

\subsection{Type IIB supergravity background}
In this subsection, we give essential details for type IIB supergravity background.
Most of the conventions in this subsection are followed from \cite{Cederwall:1996ri}.

The conventions of differential forms in target superspace
are given in accordance with the conventions given in earlier sections.
Let the 10D target superspace have mostly plus signature $(-,+,+,\cdots,+).$
A differential super $p-$form is given by
\be
C_p = \ove{p!}dZ^{M_1}dZ^{M_2}\cdots dZ^{M_p}C_{M_p\cdots M_2 M_1},
\ee
where the wedge $\w$ symbol is omitted whenever there is no risk of confusion.
The exterior derivatives and interior products act from the right.

Let coordinates on target superspace be $(Z^{M}) = (X^{m}, \th),$
where $X^m$ are $10$ bosonic coordinates,
and $\th$ are $32$ fermionic coordinates grouped into a doublet of sixteen-component Majorana-Weyl spinors
of positive chirality.
The supervielbeins are given by $(E^A) = (E^a, E^\a)$
such that
\be
E^{a} = dZ^{M}E^{a}_{M},\qquad
E^{\a} = dZ^{M}E^{\a}_{M},
\ee
where $a=0,1,2,\cdots,9$ are bosonic tangent space indices,
and $\a$ are composite indices each representing a tensor product
of a Majorana-Weyl spinor index and a doublet index.
Gamma matrices are given in the form of tensor products
\be
\G^a\otimes\s,
\ee
where $\G^a$ are gamma matrices corresponding to the Majorana-Weyl spinor,
and $\s$ is a real $2\times 2$ matrices consisting for example,
\be
\mathbbm{1}
=
\begin{pmatrix}
1 & 0\\
0 & 1
\end{pmatrix}
,\qquad
i\t_2
=
\begin{pmatrix}
0 & 1\\
-1 & 0
\end{pmatrix}
,\qquad
\t_1
=
\begin{pmatrix}
0 & 1\\
1 & 0
\end{pmatrix}
,\qquad
\t_3
=
\begin{pmatrix}
1 & 0\\
0 & -1
\end{pmatrix}
.
\ee
The Gamma matrices of the form $(\G^a\otimes\mathbbm{1})_{\a\b}, (\G^a\otimes\t_1)_{\a\b}, (\G^a\otimes\t_3)_{\a\b}$
are symmetric whereas the Gamma matrices of the form $(\G^a\otimes i\t_2)_{\a\b}$
are antisymmetric.
In practice, we will omit the notations $\otimes$ and $\mathbbm{1}.$

Apart from the supervielbeins, other target space superfields
are a dilaton $\f,$ an NS-NS $B-$field $B_2$ with field 
field strength
\be
H_3 = dB_2,
\ee
and RR C-fields $C_0, C_2, C_4,$ whose field strengths
are given respectively by
\be
R_1 = dC_0,\qquad
R_3 = dC_2 - H_3 C_0,\qquad
R_5 = dC_4 - H_3 C_2.
\ee
Type IIB supergravity theory is invariant under the supergauge transformation
\begin{align}
\d (E^a\otimes E^b\h_{ab}) = 0,\qquad
\d \f = 0,\qquad
\d B_2 = d\l_1,\nonumber\\\label{sgaugetrafo}
\d C_0 = 0,\qquad
\d C_2 = d\m_1,\qquad
\d C_4 = d\m_3 + B_2 d\m_1,
\end{align}
where $(\h_{ab}) = diag(-1,1,\cdots,1)$ is 10D Minkowski metric.

In order to prove kappa-symmetry
for a D3-brane coupled to target superspace,
the latter has to obey the constraints.
Let us quote only the relevant constraints for this purpose.
The first one is a torsion constraint given by
\be
T^a \equiv d E^a + E^b\o_b{}^a = i E^\a\G^a_{\a\b} E^\b,
\ee
where $\o_b{}^a$ are connection one-forms.
Other constraints are
\begin{align}
H &= -ie^{\f/2} E^c E^\b E^\a(\G_c\t_3)_{\a\b}
+\hlf e^{\f/2} E^c E^b  E^\a(\G_{bc}\t_3\L)_\a+\cdots,\nonumber\\
R_1 &= 2e^{-\f}E^\a(i\t_2\L)_\a+\cdots,\nonumber\\
R_3 &= ie^{-\f/2}E^c E^\b E^\a(\G_c \t_1)_{\a\b} + \hlf e^{-\f/2} E^c E^b E^\a (\G_{bc}\t_1\L)_\a + \cdots,\nonumber\\
R_5 &= \frac{i}{6}E^\b E^\a E^c E^b E^a (\G_{abc}i\t_2)_{\a\b}+\cdots,\label{fscons}
\end{align}
where $\cdots$ are terms which are irrelevant to kappa-symmetry variation,
and
\be
\L_\a = \hlf E^M_\a\pa_M\f.
\ee

\subsection{Standard sypersymmetric D3-brane action}\label{subsec:D3}
Let us now consider a D3-brane worldvolume
embedded in the target superspace.
The worldvolume coordinates are labelled by $x^\m; \m=0,1,2,3.$
Gauge field $A=dx^\m A_\m$ lives in the worldvolume.
The couple of worldvolume field with the background superfields
are explained with the help of pullbacks of the latter.
The extended field strength of $A$ is given by
\be
F = dA - B,
\ee
where $B$ here is the pullback of NS-NS $B-$field.
Note that we do not introduce an extra notation
to distinguish background superfields from their pullbacks.
We will also adopt this convention for the rest of this section.

The standard D3-brane action is given by
\be\label{D3action}
S_{D3}
=-\int d^4 x\sqrt{-\det(g_{\m\n} + e^{-\f/2}F_{\m\n})}
+\int (C_4 + F\w C_2 + \hlf F\w F\w C_0),
\ee
where $g_{\m\n} = E^a_\m E^b_\n \h_{ab},$ and $E^a_\m = E^a_M \pa_\m Z^M.$
It is invariant under the kappa-symmetry transformation define via
\be
i_\k E^{a} = \d_\k Z^{M}E^{a}_{M} = 0,\qquad
i_\k E^{\a} = \d_\k Z^{M}E^{\a}_{M} = ((\mathbbm{1}+\bar\G)\k)^\a,
\ee
in which it follows that
\be
\d_\k g_{\m\n} = 4iE^\a_{(\m}(\g_{\n)})_{\a\b}i_\k E^\b,\qquad
\d_\k\f = 2\L^\a (i_\k E)_\a,\qquad
\d_\k F = - i_\k H,
\ee
where $\g_\m$ is the pullback of $\G_a,$ i.e. $\g_\m = E^a_\m\G_a.$
Additionally,
$i_\k H, i_\k R_1(=\d_\k C_0), i_\k R_3, i_\k R_5$
are also needed and can be computed from eq. \eq{fscons}.
The matrix $\bar\G$ is computed using the standard method,
see for example \cite{Cederwall:1996pv},
and is given by
\be
\sqrt{1-\hlf e^{-\f}[F^2]-\ove{16}e^{-2\f}[F\tilde F]^2}\ \bar\G
=\bar\g\t_2
-\hlf\tilde F^{\m\n}\g_{\m\n}\t_1
+\frac{i}{4}[F\tilde F]\t_2,
\ee
where
\be
\bar\g = -\frac{i}{4!}\frac{\e^{\m\n\r\s}}{\sqrt{-g}}\g_{\m\n\r\s}.
\ee

\subsection{Covariant duality-symmetric supersymmetric D3-brane action in dual formulation}
The standard D3-brane action formulated for example in \cite{Cederwall:1996pv,Aganagic:1996pe}
does not retain $SL(2,\RR)$ symmetry of the background supergravity.
A way to remedy this is seen for example in 
\cite{Cederwall:1997ab}, in which an additional world-volume gauge field is introduced.
This allows the definition of extended field strength
of the form ``$\vec{F} = d\vec{A} - \vec{C},$''
where $\vec{C}$ is an $SL(2,\RR)$ doublet of 2-form potential in type IIB supergravity.
The doublet $\vec{F}$ is not completely independent.
Their components are related to each other
by non-linear twisted self-duality condition
which are imposed by hand, i.e. not at the action level.
This theory is verified by the construction of kappa-symmetry.

An improvement to this theory was initiated by \cite{Nurmagambetov:1998gp},
which constructed
duality-symmetric D3-brane action having non-linear twisted self-duality
condition as the field equation of worldvolume gauge fields.
The theory was constructed by dimensionally reducing
PST-covariantised bosonic M5-brane \cite{Pasti:1997gx}
which in turn made use of
an auxiliary scalar field to maintain the 
manifest diffeomorphism invariance of 6D worldvolume.
The dimensionally reduced 4D worldvolume action
retains the auxiliary scalar field and hence
the manifest worldvolume diffeomorphism invariance.
The reference \cite{Suzuki:1999aa} then extended the theory of 
\cite{Nurmagambetov:1998gp} to covariant duality-symmetric D3-brane theory coupled to
10D type IIB supergravity.
The construction was then verified by the proof of kappa-symmetry.

For us, we are also going to construct
covariant duality-symmetric D3-brane coupled to
10D type IIB supergravity, and proving kappa-symmetry.
Our construction will be differed from \cite{Suzuki:1999aa}
by the fact that we work in dual formulation of duality-symmetric theory.
Furthermore, the background supergravity used in \cite{Suzuki:1999aa}
was based on \cite{Cederwall:1997ab} where the axion and dilaton are $SL(2,\RR)/SO(2)$ coset scalars,
and the $2-$ and $4-$form potentials each transform as multiplet under $SL(2,\RR).$
However, we will keep using the background supergravity
in the form that is more standard to string theory,
i.e. the $p-$form potentials are from R-R, and NS-NS sectors.

Let us start by writing down the complete action of
the covariant duality-symmetric
action in dual formulation
for supersymmetric D3-brane in Green-Schwarz formulation.
It is given by
\be\label{actioncdsD3}
S_{cds-D3} = S_1 + S_2 + S_{WZ},
\ee
where
\be
S_1 = -\ove{2}\int\Omega(\vec F\w\cP \vec F),
\ee
\be
S_2 = -\int d^4 x\sqrt{-g}\sqrt{1-2 U + 4\lrbrk{\hlf U^2 - Y}},
\ee
\be
S_{WZ} = \int\lrbrk{ C_4 + \ove{2}\Omega(\vec F_2\w \vec C_2) - \ove{2}C_2^1\w C_2^2},
\ee
where, as usual,
\be
g_{\m\n} = E^a_\m E^b_\n \h_{ab},\qquad
F_2^i = dA_1^i - C_2^i,
\ee
\be
U = -\hlf M_{ij}f^i_\m f^j_\n g^{\m\n},\qquad
Y = \ove{8} M_{ij}f^i_\m f^j_\n g^{\n\r} M_{kl}f^k_\r f^l_\s g^{\s\m},\qquad
f^i_\m = v^\n F^i_{\n\m}.
\ee
Additionally,
\be\label{cds-D3-identify}
(M_{ij})
=
\begin{pmatrix}
e^{-\f} + C_0^2 e^\f & -C_0 e^\f\\
-C_0 e^\f & e^\f
\end{pmatrix}
,\qquad
C_2^1 = B_2,\qquad
C_2^2 = C_2.
\ee
In order to derive this action, we compare the
duality relation
\be\label{P0fulleqn}
\Gt^{\m\n} \equiv \frac{2}{\sqrt{-g}}\frac{\d S_{D3}}{\d F_{\m\n}} - \Ct^{\m\n} = \Ft^{\m\n}C_0-\frac{e^{-\f}F^{\m\n}+\ove{4}e^{-2\f}[F\Ft]\Ft^{\m\n}}{\sqrt{1-\hlf e^{-\f}[F^2]-\ove{16}e^{-2\f}[F\tilde F]^2}},
\ee
with
the non-linear twisted self-duality condition of $S_{cds-D3}:$
\be\label{fstarfulleqn}
\begin{split}
\vec f_*
&= \frac{(1-2 U)\vec f- \vec f^{\,3}}{\sqrt{1-2 U + 4\lrbrk{\hlf U^2 - Y}}}.
\end{split}
\ee
The comparison allows us to work out how $M_{ij}$
and $C_2^i$ are related to $\f, C_0, C_2, B.$
In section \ref{Bosonsample}, we have shown that
in the simple case where $M_{ij} = \d_{ij}, C_2^i = 0, \f = B_2 = C_2 = 0,$
the equations \eq{P0fulleqn} and \eq{fstarfulleqn} are equal, with the identification $F^1 = F, F^2 = G.$
Turning now to the general case, the equivalence of equations \eq{P0fulleqn} and \eq{fstarfulleqn}
gives rise to eq.\eq{cds-D3-identify}.
This completes the dictionary
between the fields in standard supersymmetric D3-brane theory
and the ones in the covariant duality-symmetric version,
as well as allowing us to
determine $S_1, S_2,$ and
\be
\int\ove{2}\Omega(\vec F_2\w \vec C_2) \subset S_{WZ}.
\ee
Next, by demanding that $S_{cds-D3}$
is invariant under the supergauge transformation eq.\eq{sgaugetrafo},
the Wess-Zumino action $S_{WZ}$ and hence
the whole $S_{cds-D3}$ can now be completely determined.

Having determined the complete $S_{cds-D3}$
and written all the fields in this theory
in terms of fields in the standard D3-brane theory,
we can now apply kappa-symmetry transformation given in subsection \ref{subsec:D3}.
For this, we also need additional condition $\d_\k a^I = 0, \d_\k F^1 = -i_\k H, \d_\k F^2 = -i_\k(R_3 - H_3 C_0).$
We also need to recompute the expression for $\bar\G,$
which can be done by following the standard method and a somewhat lengthy algebra.
In the derivation, it is convenient to make use of
$T^i$ and $S^i$ defined via
\be
\d_\k F^i_{\m\n} = 4i \bar E_{[\m}^\a(\g_{\n]} T^i)_{\a\b} i_\k E^\b + \L^\a(\g_{\m\n} S^i)_{\a\b} i_\k E^\b,
\ee
which gives
\be
T^1 = e^{\f/2}\t_3,\qquad
T^2 = {-}\lrbrk{e^{-\f/2}\t_1-e^{\f/2} C_0 \t_3},
\ee
\be
S^1 = e^{\f/2}\t_3,\qquad
S^2 = {+}\lrbrk{e^{-\f/2}\t_1+e^{\f/2} C_0 \t_3}.
\ee
They have the following properties
\be
M_{ij}i\t_2 T^j = \Omega_{ij} T^j,\qquad
-M_{ij} i \t_2 S^j = \Omega_{ij} S^j,
\ee
\be
T^iT^j
=
M^{ij} - i\t_2\Omega^{ij},\qquad
S^{[i} T^{j]} = 0.
\ee
Let us finally present the result:
\be
\bar\G
=\frac{-\bar\g\t_2
+\Omega\lrbrk{v^\m \vec f^\n i\bar\g\g_{\m\n} \vec T}
-\hlf i\Omega(\vec f_\m \vec f_\n\bar\g\g^{\m\n})}{\sqrt{1-2 U + 4\lrbrk{\hlf U^2 - Y}}}.
\ee

Let us now compare the on-shell action
of the duality-symmetric D3-brane \eq{actioncdsD3}
with the standard D3-brane \eq{D3action}.
The `on-shell' equality here means
the equality after imposing the D3-brane duality relation \eq{P0fulleqn}.
After a simple substitution, we obtain
\be\label{D3onshellcompare}
S_{cds-D3}
\stackrel{\textrm{on-shell}}{=} S_{D3} - \ove{2}\int(F_2+B_2)\w(G_2+C_2).
\ee
If we also impose the $A_\m$ field equation $\pa_\m(\sqrt{-g}(\Gt^{\m\n} + \Ct^{\m\n})) = 0$ of
D3-brane action, then the second term on the RHS
of eq.\eq{D3onshellcompare} becomes a surface term.

The on-shell agreement of duality-symmetric D3-brane action
with the standard D3-brane action was done in \cite{Khoudeir:1997sk,Berman:1998va,Nurmagambetov:1998gp}.
In these references, the D3-brane whose actions are being compared 
lives in the background with graviton, dilaton, and axion.
In our analysis, we have shown that the agreement still holds
in the case where the D3-brane
is coupled to the complete 10D type IIB supergravity background.
Although the duality-symmetric D3-brane action that we analysed
are in dual formulation, we expect that
similar result should also hold in the standard formulation
analysed in \cite{Khoudeir:1997sk,Berman:1998va,Nurmagambetov:1998gp}.

\section{Conclusion}\label{sec:conclude}
In this paper, we have discussed the construction
of covariant duality-symmetric theories in dual formulation,
and then demonstrated using an explicit example of D3-brane theory
coupled to 10D type IIB supergravity background.

The duality-symmetric theories discussed in this paper
are covariantised by using three auxiliary scalar fields.
The nonlinearisation and inclusion of interaction
with external fields are also discussed.
The constructions are closely analogous
to the duality-symmetric theories in standard formulation,
especially \cite{Pasti:2012wv} which makes use of one auxiliary scalar field
to PST-covariantise the theory.
But one of the differences is that
due to the more complicated nature of the usage of three auxiliary scalar fields
over the usage of one auxiliary scalar field in standard formulation,
we need to introduce extra tools to take care of the construction.

In order to demonstrate how to apply the construction,
we discuss concrete examples which are DBI and D3-brane.
The construction of these theories as covariant duality-symmetric theories
in standard formulation has been discussed quite extensively in the literature.
The constructions in the dual formulation
can then be done using analogous approaches,
and we have shown by explicit analysis that this is indeed the case.

The covariant duality-symmetric action in dual formulation
of DBI theory takes a DBI-like form \eq{cdsDBI}.
This feature agrees with the standard formulation counterpart.
See for example \cite{Deser:1997gq}.
We have also discussed that the DBI-like form
is a special case of BN ansatz \eq{BNansatz}
in the sense that it is a solution to the differential equation
\eq{BNcondfromZ2} with trivial complementary solution.

Finally, by comparing with the D3-brane action \cite{Cederwall:1996pv,Cederwall:1996ri},
we wrote down eq.\eq{actioncdsD3} the covariant duality-symmetric action in dual formulation
of D3-brane coupled to 10D type IIB supergravity background,
and worked out the kappa-symmetry transformation.
We have also shown that the duality relation
matches with the twisted self-duality condition.
Furthermore, the two actions are shown to agree on-shell.
These results suggest that the two actions are related.
But in order to make sure, one will need further consistency checks.
One of important checks, which will be investigated as a future work,
would be to start from constructing
Hamiltonian for the action \eq{actioncdsD3}.
This would then allow one to infer important physical consequences,
for example to check whether energy of BIon solution \cite{Callan:1997kz,Gibbons:1997xz}
of the theory \eq{actioncdsD3} would agree with the one from standard D3-brane action
\eq{D3action}.

It would be natural to address the constructions
of other covariant nonlinear duality-symmetric theories in dual formulation
as future works. For example, one may wish to study
the actions which include derivatives
of field strengths, or some supersymmetric actions.
Some of these theories were
discussed in standard formulation in \cite{Bossard:2011ij,Carrasco:2011jv}.
So one will need to construct these theories
in the context of dual formulation and compare with the results
in standard formulation.

Another possible direction is based on the fact that
this paper also makes use of covariantisation using more than one auxiliary field.
One might try to construct duality-symmetric theories
in higher dimensions in formulations which allow
covariantisation using more than one auxiliary field.

\subsection*{Acknowledgements}
The author is very grateful to Dmitri Sorokin for various helpful remarks and comments on the manuscript
and to Sheng-Lan Ko for interests, discussions, and comments on the manuscript.

This work is supported by
the Office of the Higher Education Commission (OHEC)
and the Thailand Research Fund (TRF)
through grant number MRG6080136.
The author would like to thank his mentor, Burin Gumjudpai,
for helps and advices especially on the
management of this grant.


\providecommand{\href}[2]{#2}\begingroup\raggedright\endgroup

\end{document}